\documentclass[useAMS,usenatbib]{mn2e}
\usepackage{graphicx}
\usepackage{amsmath}
\usepackage{makecell}
\usepackage{gensymb}
\usepackage{macros}
\usepackage[usenames,dvipsnames,svgnames,table]{xcolor}
\usepackage[draft]{hyperref}
\definecolor{darkblue}{rgb}{0.0,0.0,0.3}
\hypersetup{colorlinks,breaklinks,
            linkcolor=darkblue,urlcolor=darkblue,
            anchorcolor=darkblue,citecolor=darkblue}
\usepackage{txfonts}
\DeclareSymbolFont{cmletters}{OML}{cmm}{m}{it}
\DeclareMathSymbol{v}{\mathalpha}{cmletters}{"76}

\newcommand{\hammer}{{\sc h-amr}}
\newcommand{\HARM}{{\sc harm}}
\newcommand{\harm}{{\sc harm}}
\newcommand{\RedeclareMathOperator}[2]{\renewcommand{#1}{}\let#1\relax\DeclareMathOperator{#1}{#2}}

\newcommand\simless\lesssim
\newcommand\simgreat\gtrsim

\title[]{Large-Scale Poloidal Magnetic Field Dynamo Leads to Powerful Jets in GRMHD Simulations of Black Hole Accretion with Toroidal Field}
\author[Liska, Tchekhovskoy, Quataert]{M. Liska$^1$\thanks{E-mail: matthewliska92@gmail.com}, A. Tchekhovskoy$^{2,3}$, E. Quataert$^{3}$\\
$^{1}$Anton Pannekoek Institute for Astronomy, University of Amsterdam, Science Park 904, 1098 XH Amsterdam, The Netherlands\\
$^{2}$Center for Interdisciplinary Exploration \& Research in Astrophysics (CIERA),
Physics \& Astronomy, Northwestern University, Evanston, IL 60208, USA\\
$^{3}$Departments of Astronomy and Physics, Theoretical Astrophysics Center, University of California Berkeley, Berkeley, CA 94720-3411, USA
}

\begin{document}

\date{Accepted. Received; in original form}
\pagerange{\pageref{firstpage}--\pageref{lastpage}} \pubyear{2018}

\maketitle
\label{firstpage}

\begin{abstract}
Accreting black holes (BHs) launch relativistic collimated jets, across many decades in luminosity and mass, suggesting the jet launching mechanism is universal, robust and scale-free.  Theoretical models and general relativistic magnetohydrodynamic (GRMHD) simulations indicate that the key jet-making ingredient is large-scale poloidal magnetic flux. However, its origin is uncertain, and it is unknown if it can be generated in situ or dragged inward from the ambient medium. Here, we use the GPU-accelerated GRMHD code \hammer{} to study global 3D BH accretion at unusually high resolutions more typical of local shearing box simulations. We demonstrate that turbulence in a radially-extended accretion disc can generate large-scale poloidal magnetic flux in situ, even when starting from a purely toroidal magnetic field.  The flux accumulates around the BH till it becomes dynamically-important, leads to a magnetically arrested disc (MAD), and launches relativistic jets that are more powerful than the accretion flow. The jet power exceeds that of previous GRMHD toroidal field simulations by a factor of 10,000. The jets do not show significant kink or pinch instabilities, accelerate to $\gamma \sim 10$ over 3 decades in distance, and follow a collimation profile similar to the observed M87 jet.
\end{abstract}

\begin{keywords}
accretion, accretion discs -- black hole physics -- %
MHD -- galaxies: jets -- methods: numerical
\end{keywords}

\section{Introduction} 
BHs can launch relativistic jets by converting BH spin energy into Poynting flux \citep{bz77}. %
The ratio of jet and accretion powers, or jet efficiency, is
maximum when the BH is both rapidly spinning and has
accumulated a substantial amount of large-scale poloidal (i.e., confined to a meridional, $R{-}z$ plane) magnetic flux
(see e.g. \citealt{2001MNRAS.326L..41K,tchekhovskoy10}). 
Given enough poloidal magnetic flux, a magnetically
arrested disc (MAD) can form
(e.g. \citealt{Narayan2003}), with jet
efficiency exceeding $100\%$ for thick 
(\citealt{Tchekhovskoy2011,Tchekhovskoy2012,Mckinney2012}) and reaching
$50\%$ for thin discs (with an aspect ratio $h/r=0.03$;
\citealt{Liska2018C}).  %

\begin{figure}
  \begin{center}
    \includegraphics[width=\columnwidth,trim=0mm 1mm 0mm 4mm,clip]{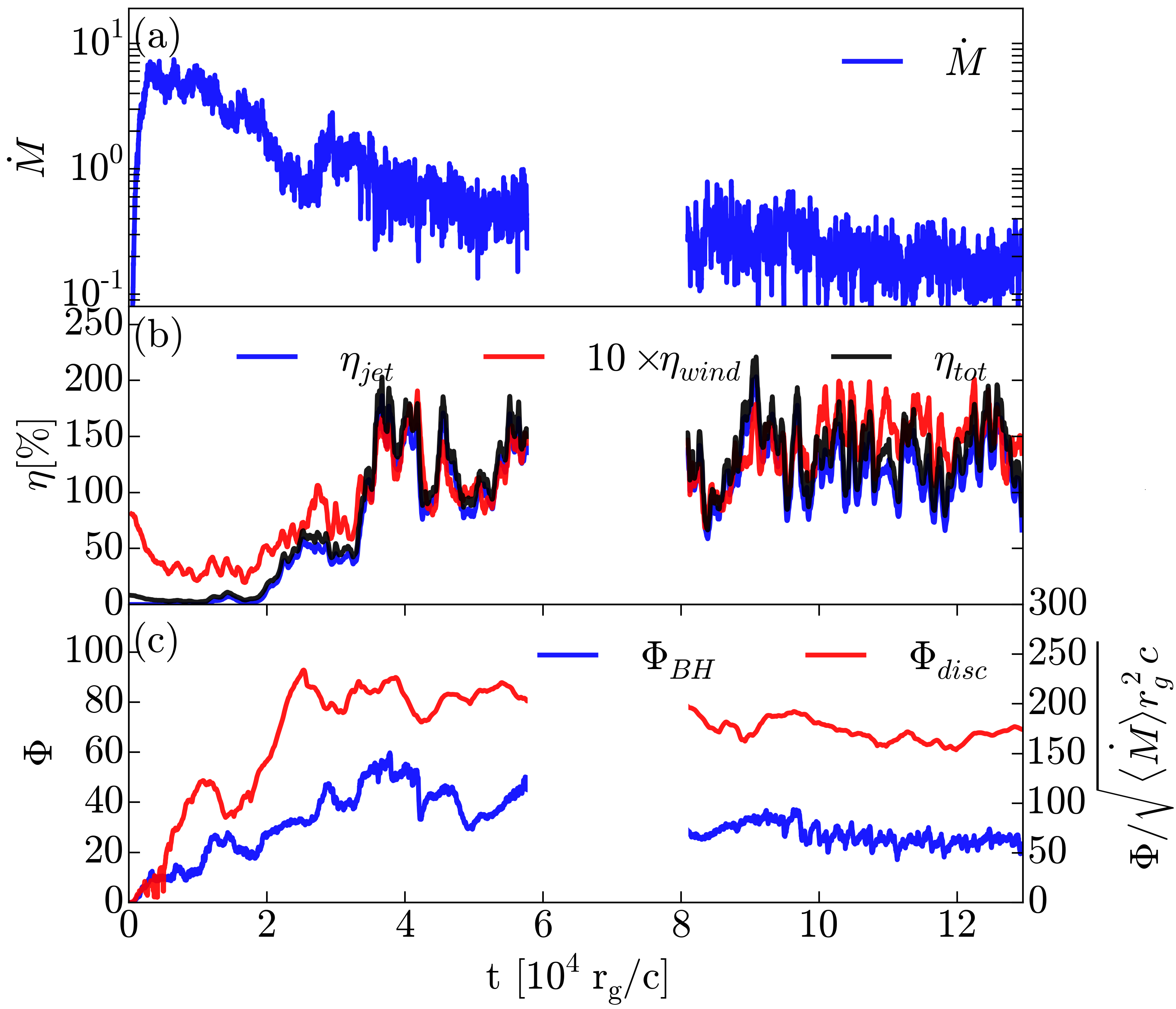}
  \end{center}
  \caption{Time evolution of various quantities {(data between $t=5.7 \times 10^4$ and $t=8.2 \times 10^4 r_g/c$ is missing)}. (a) The mass accretion rate $\dot{M}$ peaks at early times and gradually decreases; (b) The jets start out much weaker than the disc wind, $\eta_{\rm jet}\ll\eta_{\rm wind}$. However, the two become comparable at $t=(1.2{-}1.8)\times 10^4r_g$, and at later times the situation reverses. Eventually, $\eta_{\rm jet}$ exceeds $100\%$, i.e., jet power exceeds the accretion power. (c) This increase in jet power comes from the increase in the strength of BH poloidal magnetic flux, $\Phi_{\rm BH}$, due to the BH accreting dynamo-generated poloidal magnetic flux. Eventually, $\Phi_{\rm BH}$ exceeds a critical dimensionless value, $\Phi/(\langle\dot{M}\rangle r_g^2 c)^{1/2}\sim 50$, as seen from the right axis that shows the BH flux normalized by the late-time value of $\langle \dot M\rangle \approx 0.2$: $\Phi_{\rm BH}$ becomes dynamically important and leads to a MAD. A reservoir of positive poloidal magnetic flux $\Phi_{\rm disc}$ remains in the disc and may reach the BH at later times.}
\label{fig:timeplot}
\end{figure}

One way of obtaining the large-scale poloidal magnetic flux near the BH is advecting it from large radii. While thick discs can do this over short distances (as confirmed by 3D GRMHD simulations, e.g., \citealt{2006ApJ...641..103H,Tchekhovskoy2011}), it is unclear whether they can do this over 5--6 orders of magnitude in distance, from the ambient medium all the way down to the BH. This is particularly uncertain given that in many systems the accretion discs at large radii are expected to cool radiatively and become thin.
In such discs the poloidal magnetic flux may diffuse out faster than it can be advected inwards (\citealt{Lubow1994}; however, see \citealt{Rothstein2008,Giulet2012,Giulet2013}).

Large-scale poloidal magnetic flux on the BH can also form in situ through a turbulent dynamo \citep{Brandenburg1995} powered by the magnetorotational instability (MRI, see \citealt{Balbus1991}).  Local shearing box studies found that the dynamo can produce radial and toroidal magnetic fluxes on the scale of the box (e.g., \citealt{Brandenburg1995,1996ApJ...463..656S, 2008A&A...488..451L,Davis2010, Simon2012,Salvesen2016,2016MNRAS.456.2273S,2017ApJ...840....6R}).
However, persistent jets lasting an accretion time require poloidal magnetic flux on a much larger scale.
Additionally, the often used quasi-periodic boundary conditions in the horizontal direction imply that the net vertical magnetic flux through the box cannot change in time: in fact, it is a crucial externally-imposed parameter. Shearing box simulations without net flux do not appear to generate poloidal magnetic flux that affects the turbulence in the same way the net flux does \citep{2007ApJ...668L..51P,Bai2012,Salvesen2016,2016MNRAS.460.3488S,2016MNRAS.462..818B}.

Free from these limitations, global GRMHD simulations are particularly attractive for studying the formation of large-scale poloidal magnetic flux and the associated jets and outflows. Most global simulations have focused on the initial seed poloidal magnetic flux, in the form of one or several poloidal magnetic field loops; and those show no signs of a large-scale poloidal magnetic flux dynamo (e.g., \citealt{Mckinney2006,2006ApJ...641..103H,2008ApJ...687L..25S,Noble2009,Penna2010,2012MNRAS.426.3241N}). \citet{2008ApJ...678.1180B} found that jets formed only for initial conditions with net poloidal magnetic flux, and no jets formed for a purely toroidal initial magnetic flux. For a similar toroidal magnetic field initial condition, but for a larger initial torus, \citet{Mckinney2012} found short-lived jets with duration $\lesssim 12 M_{\rm BH}/(10^8M_\odot)$ days and a low duty cycle, $\sim2$\%, where $M_{\rm BH}$ is BH mass. Their low time-average efficiency, $\lesssim0.01\%$, implied that 
such weak jets would disrupt easily through the kink instability (\citealt{Tchekhovskoy2016,Bromberg2016}). This low efficiency also appears insufficient to account for feedback from AGN jets on kiloparsec scales (e.g. \citealt{Fabian2012}) and for the substantial jet power inferred in AGN jets (see, e.g., \citealt{Prieto2016} for M87, \citealt{2015MNRAS.449..316N} for low-luminosity AGN, and \citealt{Ghissellini2014} for blazars). Thus, there appears to be a serious mismatch between theory and observations due to the inability of GRMHD simulations to generate sufficient large-scale poloidal magnetic flux starting without one initially.

Yet, toroidal magnetic flux is a natural starting point for accretion discs in various contexts. %
In compact object mergers, the orbital shear is expected to produce a toroidally-dominated magnetic field geometry. In X-ray binaries, the stream overflowing the Roche lobe (or wind from the companion star) would stretch out in the toroidal and radial directions as it feeds the outer disc, and the disc shear would then substantially amplify the toroidal component. 
Similarly, the tidal debris stream feeding the supermassive BH during a tidal disruption event (TDE) is also expected to lead to a toroidally-dominated magnetic field.
{While AGN appear to have more than sufficient large-scale magnetic flux in the interstellar medium, it is unclear whether their thin accretion disks can drag it to the BH. Thus, it is important to understand whether accretion discs can produce large-scale poloidal magnetic flux on their own.}

This motivates our study of BH accretion seeded with purely a toroidal magnetic flux. In Sec.{}~\ref{sec:numerical-models} we describe the numerical setup, in Sec.{}~\ref{sec:Results} we present our results, and in Sec.{}~\ref{sec:Conclusion} we conclude.

\begin{figure*}
  \begin{center}
    \includegraphics[width=\linewidth,trim=0mm 2mm 0mm 0mm,clip]{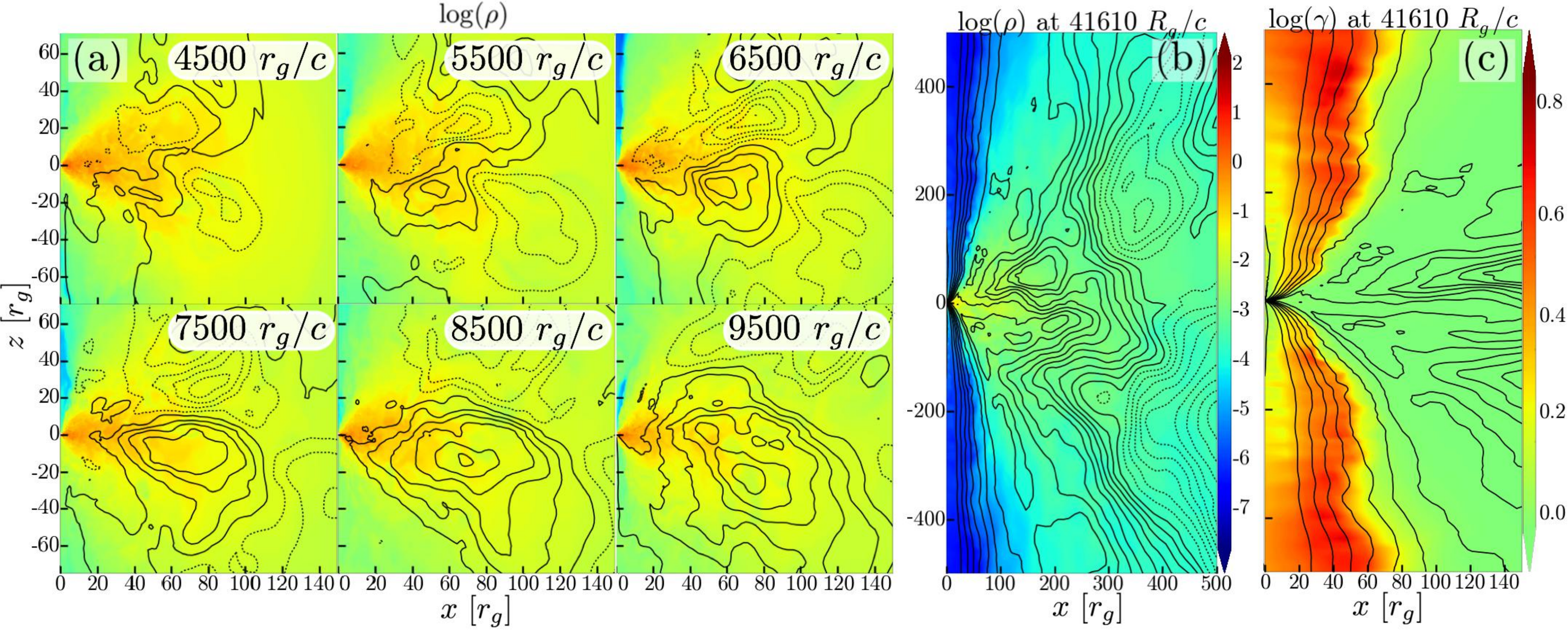}
    \end{center}
\caption{(a) Meridional slices through the simulation at the approximate times shown, illustrating the development of large scale poloidal flux loops (black lines) of size comparable to the disc thickness, plotted over the density distribution (shown in colour; see the colour bar in panel b). The loops form slightly offset from the equator and buoyantly rise away from it, as expected in an $\alpha{-}\Omega$ dynamo. %
  Initially, several poloidal field loops of positive (solid lines) and negative (dotted lines) polarity stochastically form. However, most of them get expelled, and one largest, lucky, loop takes over. See SI and \href{https://www.youtube.com/channel/UCB2BslhkPMTkIicfoIBG0mQ}{this link} for a movie. (b) A snapshot at $t=4.2\times10^4r_g/c$ reveals two large-scale dynamo-generated poloidal magnetic flux loops: their size, $l\gtrsim100r_g$, vastly exceeds that of the event horizon, and the loops present themselves to the BH as large-scale poloidal flux. (c) The colour map of the Lorentz factor, $\gamma$, shows that this flux leads to the launching of relativistic jets with a typical spine-sheath structure. {The movie (see SI) shows no obvious signs of global kink or pinch modes in the jets, which reach $\gamma \sim 5$ at $z \lesssim 500 r_g$. We see no signs of such modes out to $z\sim2000r_g$ by which the jets reach $\gamma\sim10$ (not shown due to space constraints).}
    }
\label{fig:contourplot}
\end{figure*}

\section{Numerical Approach and Problem Setup}
\label{sec:numerical-models}
We use the \hammer{} code \citep{Liska2018A,2019arXiv191210192L} that evolves the GRMHD equations of motion (\citealt{Gammie2003}) on a spherical polar--like grid in Kerr-Schild coordinates, using PPM spatial reconstruction  \citep{1984JCoPh..54..174C} and second order time-stepping. \hammer{} includes GPU acceleration and advanced features, such as adaptive mesh refinement (AMR) and local adaptive timestepping.

We start with a BH of dimensionless spin $a = 0.9$ surrounded by an equilibrium hydrodynamic torus with a sub-Keplerian angular momentum profile, $\ell \propto r^{1/4}$, inner edge at $r_{\rm in} = 6r_g$, density maximum at $r_{\rm max} = 13.792 r_g$, and outer edge at $r_{\rm out}=4\times10^4r_g$ \citep{1985ApJ...288....1C,2003ApJ...589..458D}; $r_g = GM_{\rm BH}/c^2$ is the gravitational radius. The torus aspect ratio ranges from $h/r=0.2$ at $r_{\rm max}$ to $0.5$ at $r_{\rm out}$. We insert into the torus toroidal magnetic field with a uniform plasma $\beta = p_{\rm gas}/p_{\rm mag}=5$ and add random 5\%-level perturbations to $p_{\rm gas}$ to seed the non-axisymmetric MRI.

Operating in spherical polar coordinates, with a lo\-ga\-rith\-mi\-cal\-ly-spaced $r$-grid and uniform $\theta$- and $\phi$-grids, we use transmissive boundary conditions (BCs) at the poles, $\sin\theta=0$ (see the supplementary information [SI] in \citealt{Liska2018A}), periodic BCs in the $\phi$-direction, and absorbing BCs at the inner and outer radial boundaries, located just inside of the event horizon and at $r=10^5 r_g$, respectively (thus, both radial boundaries are causally disconnected from the accretion flow). We use a resolution of $N_r\times N_\theta\times N_\phi = 1872\times624\times1024$, resulting in a total of $1$ billion cells. This results in $70{-}90$ cells per disc scale height, $h/r\approx0.35{-}0.45$. Such high resolutions are typically reserved for local shearing box simulations. To increase the timestep and maintain a near-unity cell aspect ratio everywhere, we reduce the $\phi$-resolution near the pole (at $\sin\theta<0.5$, using 4 AMR levels, from $N_\phi = 1024$ at equator to $N_\phi = 128$ at the poles; see also \citealt{2019arXiv191210192L}). We carry the simulation out to $t_{\rm F} \approx 1.3\times 10^5r_g/c$.

\begin{figure}
\begin{center}
\includegraphics[width=1\columnwidth,trim=0mm 6mm 0mm 29mm,clip]{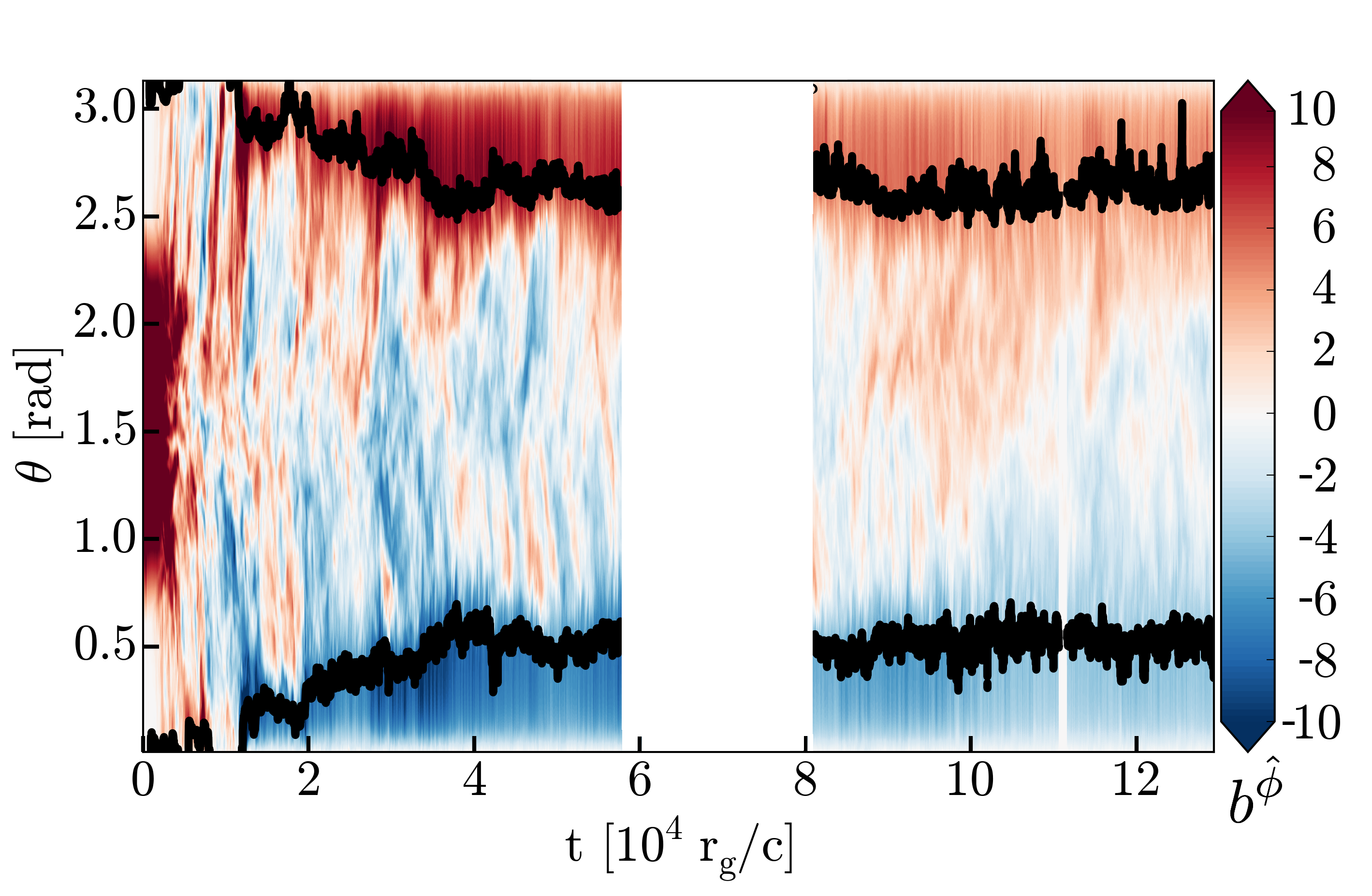}
\end{center}
\caption{The space-time diagram of toroidally averaged rest-frame magnetic field $b^{\hat{\phi}}$ at $r = 40$ $r_g$ shows {an irregular butterfly pattern, indicating sporadic dynamo activity characteristic of thick, sub-Keplerian, or strongly magnetized disks (see Sec.~\ref{sec:Results}).} Black lines track the disk-jet boundary. 
}
\label{fig:butterfly}
\end{figure}

\section{Results}
\label{sec:Results}
Figure~\ref{fig:timeplot}(a) shows that after peaking, mass accretion rate remains approximately constant at $t\lesssim 10^4r_g/c$.  %
At this time, the only outflow present is a sub-relativistic wind with
energy efficiency $\eta_{\rm wind} \approx 5\%$ (Fig.~\ref{fig:timeplot}b).
Poloidal magnetic flux on
the BH, $\Phi_{BH} = 0.5\int_{r=r_{\rm H}}\left|B^r\right| dA_{\theta\phi}$, grows from $0$ to $20$, as shown by the blue line (Fig.~\ref{fig:timeplot}c). Here, the integral is over both hemispheres of the event horizon, $r_{\rm H}=r_g[1+(1-a^2)^{1/2}]$, and the factor of $0.5$ converts it to one hemisphere \citep{Tchekhovskoy2011}.
The reservoir of positive poloidal magnetic flux in the disc, $\Phi_{\rm disc}=\max_{r} \Phi_{p}(r)$ with $\Phi_{p}(r)=\max_{\theta}\int_0^\theta B^r dA_{\theta\phi}$, shown with the red line, also keeps growing, pointing to a large-scale dynamo activity in the disc. Here, $\max_r$ and $\max_\theta$ refer to the maxima taken over $r$ and $\theta$ coordinates, respectively.

Figure~\ref{fig:contourplot}(a) shows a time sequence illustrating the generation of poloidal magnetic flux loops by the MHD turbulence (see \href{https://www.youtube.com/channel/UCB2BslhkPMTkIicfoIBG0mQ}{the movie} in the SI): several loops form just outside the equatorial plane, grow in strength, and buoyantly rise away from the equator. This process is stochastic: one of the loops ends up taking over the inner $100r_g$ of the disc with the others getting expelled in outflows. 
This is consistent with the $\alpha{-}\Omega$ large-scale poloidal magnetic flux dynamo \citep{1955ApJ...122..293P,1978mfge.book.....M}: a toroidal magnetic field loop undergoes Parker instability, buoyantly rises, and the Coriolis force twists it into a poloidal magnetic field loop. This way, the $\alpha$-effect can convert toroidal into poloidal flux. The $\Omega$-effect then does the opposite, shearing out this freshly-generated poloidal magnetic flux loop into toroidal magnetic flux, and thereby completing the positive feedback cycle. This is a possible mechanism for both the initial formation of the poloidal magnetic flux loops and their subsequent runaway growth in strength and size, as seen here.

This picture is consistent with the butterfly diagram in Fig.~\ref{fig:butterfly}: patches of toroidal magnetic field, $b^{\hat\phi}$, rise with alternating signs away from the equator. However, our dynamo is rather sporadic and irregular, reminiscent of
lower plasma $\beta$ (e.g. \citealt{Bai2012,Salvesen2016}) and sub-Keplerian \citep{2015MNRAS.446.2102N} shearing box simulations, and global simulations of very thick discs (\citealt{2018ApJ...861...24H,2019MNRAS.482..848D}) that show similar irregularity and even complete absence of sign flips. Global simulations at high $\beta$ tend show a more regular butterfly diagram  \citep{Shi2010,Oneill2011,Flock2012,Beckwith2011,Simon2011,Simon2012,Jiang2017,Siegel2017}. 

\begin{figure}
  \begin{center}
    \includegraphics[width=\linewidth,trim=0mm 2mm 0mm 0mm,clip]{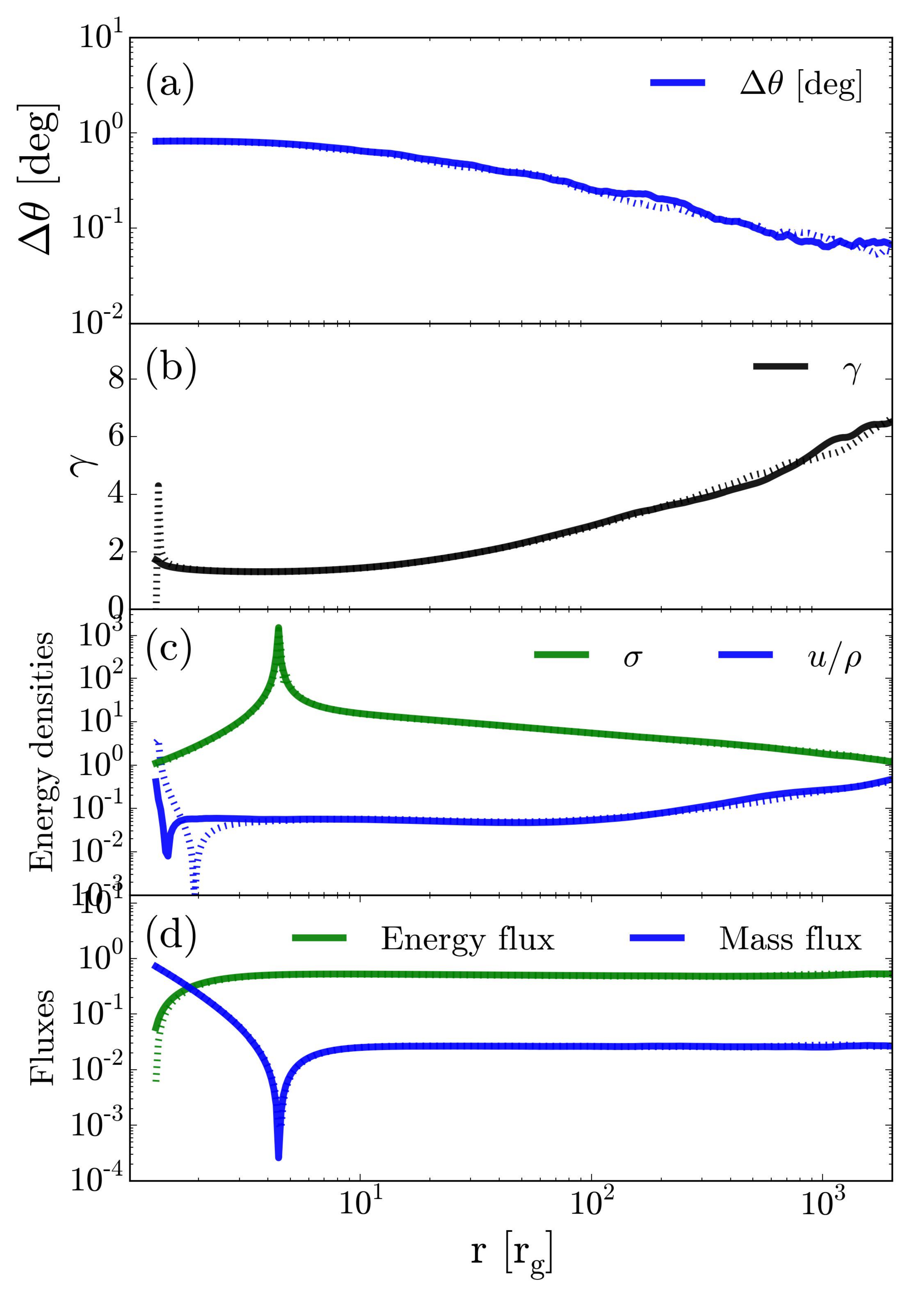}
    \end{center}
\caption{Radial profiles of the upper (dashed line) and lower (solid line) jet at $t=3.75\times10^4 r_g/c$ averaged over $\Delta t = 5\times10^3 r_g/c$. (a) The half opening angle of the jet ($\Delta \theta$) decreases as the jet is collimated by the disk wind. (b) The jet readily accelerates to $\gamma \sim 6$ (weighted by the energy flux) over $2000 r_g$. Parts of the jet reach higher Lorentz factors (see Fig.~\ref{fig:jet_zoomout}). (c) As the jet accelerates the magnetization ($\sigma=b^2/\rho$) drops. Part of this magnetic energy is converted into kinetic energy and part of it heats up the jet leading to a rise in the specific internal energy ($u/\rho$). (d) The mass and energy fluxes integrated over the jet's opening angle remain (almost perfectly) conserved over $1000 r_g$. }
\label{fig:jetplot}
\end{figure}

Figure~\ref{fig:timeplot}(c), right axis, shows that the magnetic flux grows until the critical value $\Phi/(\langle\dot M\rangle r_g^2 c)^{1/2} \approx 50$ \citep[][]{Tchekhovskoy2011} at which the BH flux becomes dynamically-important, obstructs accretion, and leads to a MAD. Figure~\ref{fig:timeplot}(b) shows that jets reach $\eta_{\rm jet} \approx 150\%$, comparable to or exceeding $100$\%: a tell-tale signature of the MAD state. Figure~\ref{fig:jetplot}(a),(b) shows that the jets collimate to small aspect ratios, $R/z\approx0.3$, $0.12$, $0.08$, and accelerate to relativistic Lorentz factors, $\gamma\sim3$, $5$, $10$ at $z/r_g = 100$, $500$, $2000$, respectively, similar to the observed M87 galaxy jet (see Appendix~\ref{sec:large-scale-jet}; \citealt{2013ApJ...775..118N,2016A&A...595A..54M}; see also \citealt{2019arXiv190403243C}). During this process the jet converts magnetic energy into kinetic energy and heat, while the total mass and energy fluxes remain conserved, as seen in Fig.~\ref{fig:jetplot}.

Figure~\ref{fig:radplot}(a) shows that whereas at early times the magnetic pressure in the disc is mostly subdominant, at later times it comes close to equipartition as characteristic of MADs \citep{Mckinney2012}. This leads to relatively high $\alpha$-viscosity in the disc, with Maxwell and Reynolds stress contributions of $\alpha_M=b_{\hat r} b_{\hat\phi}/(p_g+p_b) \simeq0.1$ and $\alpha_R=\rho u_{\hat r} u_{\hat\phi}/(p_g+p_b)\simeq0.01$, respectively; here the hats indicate the physical components. Such high Maxwell stresses are atypical and were only found in the presence of large-scale poloidal magnetic flux threading the disc \citep{Mckinney2012,Bai2012,Salvesen2016}: indeed, while $40\%$ of the poloidal magnetic flux reaches the BH, the disc retains the rest; see Figs.~\ref{fig:contourplot}(b),(c) and \ref{fig:radplot}(c). 
{If such in-situ generation of equipartition ($\beta \sim 1$) magnetic fields from sub-equipartition toroidal fields carries over to thin disks, this may stabilize them against the viscous-thermal instability (\citealt{Begelman2007}; see also \citealt{2019arXiv190401674J,2016MNRAS.459.4397S}).} 
Interestingly, our recent simulation of a thin disk (with $h/r=0.02$) initially threaded with a purely toroidal magnetic flux, did not show any signs of relativistic jets \citep{2019arXiv191210192L}. This suggests that the large-scale poloidal magnetic flux generation proceeds more efficiently for thick discs than for their thin counterparts.

\section{Discussion and Conclusions}
\label{sec:Conclusion}
Using global GRMHD simulations at one of the highest resolutions to date, we demonstrate for the first time that a large-scale poloidal magnetic flux dynamo operates in BH accretion discs.  Poloidal field loops form in situ, of size $l\lesssim h\sim r$, slightly offset from the equator, and tend to rise buoyantly (Fig.~\ref{fig:contourplot}a). The formation mechanism is consistent with the $\alpha{-}\Omega$ dynamo, which relies on the buoyancy and Coriolis forces to convert toroidal into poloidal magnetic flux ($\alpha$-effect), and on the disc shear to convert poloidal into toroidal flux ($\Omega$-effect). With most of the loops expelled, the lucky remaining loop structure resides at $r\gtrsim 200r_g$ (Fig.~\ref{fig:contourplot}b). It presents itself to the BH as a \emph{large-scale poloidal magnetic flux}, whose scale exceeds the local radius by 2 orders of magnitude.
The flux accumulates around the BH until it becomes dynamically-important, and leads to a MAD and mag\-ne\-ti\-cal\-ly-laun\-ched jets of constant magnetic polarity whose power exceeds the accretion power (Fig.~\ref{fig:timeplot}b).

With the grid totalling $1$ billion cells and effective resolution of $N_r\times N_\theta\times N_\phi = 1872\times624\times1024$ cells, this is the highest resolution simulation of MADs. The much increased resolution -- an increase in the number of cells by factors of approximately $200$ compared to \citet{Tchekhovskoy2011} and $60$ compared to \citet{2019ApJ...874..168W} -- allows us to check whether the properties of MADs are sensitive to the changes in the resolution. We note that this unusually high resolution is combined with the extremely long duration of $t_{\rm F} \approx 1.3\times10^5r_g/c$, which allows us to ensure that the values of efficiency and magnetic flux have reached their asymptotic values and the simulation is firmly in the MAD regime. The time-average values of efficiency, $\eta_{\rm tot}\approx 150\%$ (Fig.~\ref{fig:timeplot}b), and dimensionless magnetic flux, $\phi_{\rm BH}\approx 50$ (Fig.~\ref{fig:timeplot}c), are in agreement with previous findings for MADs at this thickness ($h/r\sim 0.35{-}0.45$). This suggests that the time-average results are numerically converged. Note that MADs are typically simulated by starting with initial conditions that contain a large amount of poloidal magnetic flux. In contrast, in this work, we started with a purely toroidal magnetic flux. That we find quantitively similar results suggests that so long as the system ends up in a MAD state, the outcome is not sensitive to the path the system took to get there or the exact initial conditions it started with.

\begin{figure}
\begin{center}
\includegraphics[width=0.9\columnwidth,trim=0mm 20mm 0mm 7mm,clip]{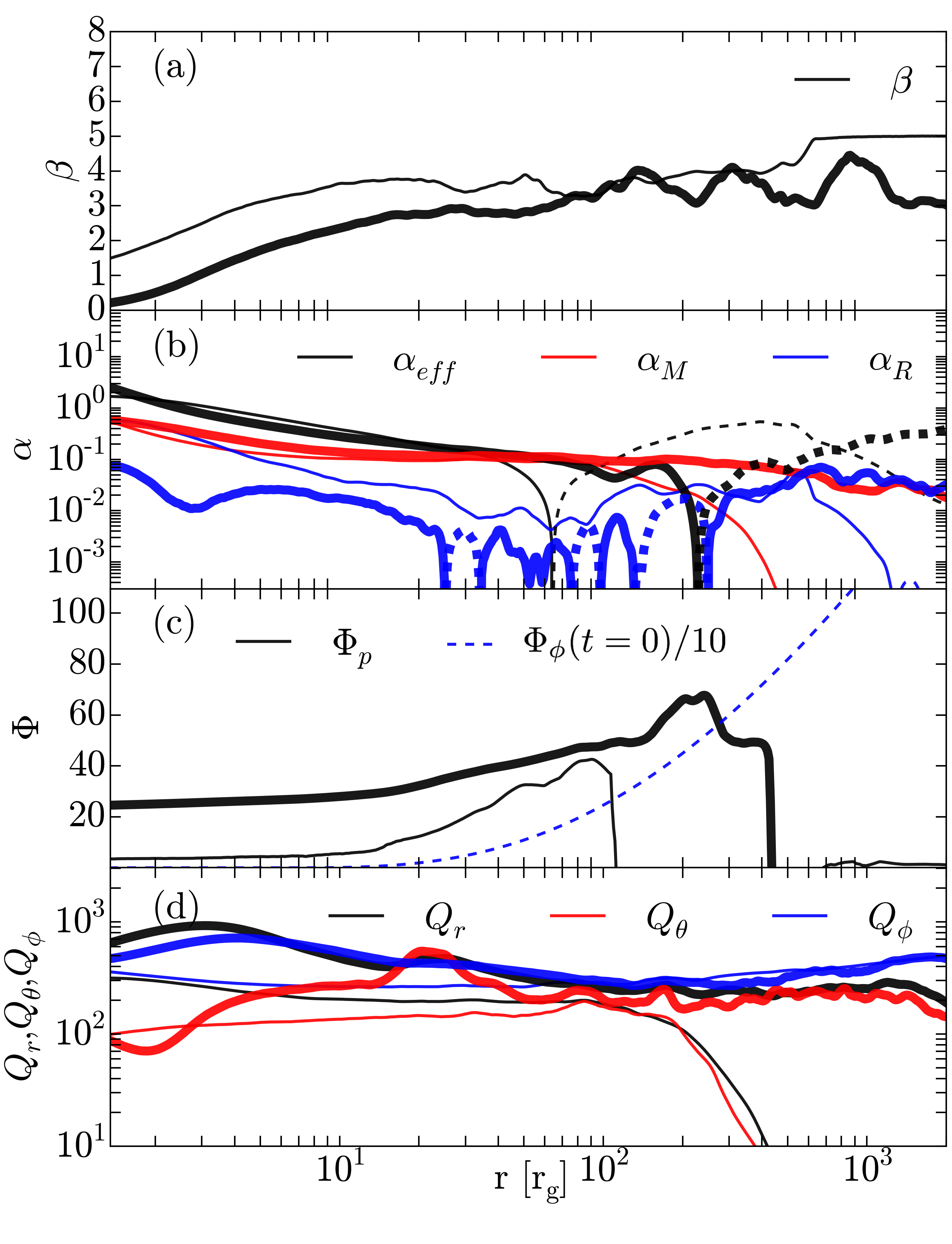}
\end{center}
\caption{Radial profiles at $t=10^4r_g/c$ (thin lines) and $t=1.3\times10^5$ $r_g/c$ (thick lines), averaged over $\Delta t = 10^3r_g/c$. (a) While at early times the magnetic pressure is subdominant ($\beta \sim$ few), at late times the magnetic pressure reaches equipartition ($\beta\lesssim1$), as the dynamo-generated poloidal magnetic flux increases in strength. (b) The effective viscosity, $\alpha_{\rm eff} = -v_{\hat r}v_{\hat \phi}/c_{\rm s}^2$, and the contributions to viscosity of Maxwell ($\alpha_{M}$) and Reynolds ($\alpha_R$) stresses. Solid curves show positive and dashed negative values. Maxwell stress dominates over Reynolds stress. Near the BH we have $\alpha_{\rm eff}>\alpha_{M}$, likely due to plunging of the gas into the BH. Stagnation points in the flow, where $\alpha_{\rm eff} = 0$, are favorable locations for dynamo operation (see Sec.~\ref{sec:Conclusion}). (c) The disc has a large reservoir of positive polarity magnetic flux $\Phi_p$, whose peak moves outward with time. 
  {Blue dashed line shows that the initial toroidal magnetic flux $\Phi_{\phi}(t=0)$ (scaled by 0.1x) exceeds by a factor of $\sim7$ the late-time poloidal flux} at $200r_g$, suggesting that even more poloidal magnetic flux may be generated at times beyond those simulated. (d) High values of quality factors, $Q_\#$, the number of cells per MRI wavelength in direction $\#$, show that the turbulence is very well resolved in all three dimensions.
}
\label{fig:radplot}
\end{figure}

We initialized our simulations with a relatively strong toroidal magnetic field, with plasma $\beta = 5$, 
to ensure that they resolve the MRI due to both the initial toroidal and dynamo-generated poloidal magnetic fields: we found that the dynamo did not operate in the simulations that formally resolved the former but not the latter. Figure~\ref{fig:radplot}(d) shows that the MRI is well-resolved, 
with $Q_{r,\theta}\gtrsim100$ cells per MRI wavelength in the $r$-, and $\theta$- and $Q_{\phi}\gtrsim200$ in the $\phi-$direction.
Using the same physical setup, same effective $\theta$-resolution near the equator, and 4 times lower $\phi$-resolution, resulted in twice as low outflow energy efficiency of $\eta_{\rm tot} \simeq 25$\% at $t = 2.5\times10^4r_g/c$ (obtained with the HARM code at a resolution of $288\times128\times128$, with the $\theta$-grid focused on the equator and using a toroidal wedge, $\Delta\phi = \pi$; see Appendix~\ref{sec:descr-low-resol}). This suggests that for the toroidal flux to generate large-scale poloidal flux it is crucial that the MHD turbulence is well-resolved.

Figure~\ref{fig:radplot}(c) shows that the poloidal flux produced in our simulation makes up $\lesssim 15\%$ of the initial toroidal flux, $\Phi_\phi(r)=\int_{0}^{r} B^{\phi} dA_{r\theta}$, at $r\lesssim 200 r_g$; here, the integral is in the $r$- and $\theta-$directions spanning the full range of $\theta\in[0,\pi]$. Thus, large-scale toroidal flux might be a prerequisite for the large-scale poloidal flux dynamo to operate. 
Assuming that the dynamo converts a fixed fraction of {the initial} toroidal flux into poloidal flux, it will take longer to generate the same poloidal flux for a weaker initial toroidal flux or, equivalently, higher value of plasma $\beta$. If the dynamo-generated poloidal magnetic flux is limited by the time available for the dynamo to operate, stagnation points in the disc -- where the gas lingers instead of falling in or flying out and where the effective viscosity $\alpha_{\rm eff}= -v_{\hat r}v_{\hat \phi}/c_{\rm s}^2$ vanishes -- can become centers of poloidal flux generation. Figure~\ref{fig:radplot}(b) shows that the radius of the stagnation point changes very slowly: from $65r_g$ at $t = 10^4r_g/c$ to $230r_g$ at $t = 1.3\times10^5r_g/c$. The gas flow is directed away from the stagnation point in all directions, and this expanding flow pattern assists the dynamo not only in inflating the poloidal flux loop but also in trapping it at the stagnation point. As seen in Figure~\ref{fig:contourplot}(a), this trapping might be responsible for one lucky loop getting pinned down at the stagnation point, outgrowing the rest of the loops, and dominating the long-term evolution of the system. In a similar way, even for an initial small-scale toroidal magnetic field, a stagnation point may trap a poloidal magnetic flux loop and inflate it to large scales, thereby producing large-scale poloidal magnetic flux.
Figure~\ref{fig:radplot}(c) shows that in our simulations the peak of the poloidal flux grows in amplitude and moves out to a larger distance, loosely following the movement of the stagnation point. This suggests that the dynamo might indeed benefit from the presence of stagnation points in the flow. {Since the stagnation point in externally fed accretion discs in XRBs and AGN would form around the disc's circularization radius located order(s) of magnitude further away from the BH than in this work, the buildup of poloidal flux on the BH would take much longer than presented in this work.}

The lack of large-scale poloidal flux generation in global toroidal field GRMHD simulations till now might stem from a lower field strength considered, lack of stagnation points suitably located in the flow, a limited radial range of the initial toroidal magnetic flux distribution \citep{2008ApJ...678.1180B} or a high radial inflow velocity (due to large disc thickness) that may not give the poloidal field loops enough time to grow \citep{Mckinney2012}. In addition, our work seems consistent with \citet{Fragile2017}, which for a strong $\beta \sim 0.1$ magnetic field did not find any generation of large scale poloidal magnetic flux loops on a rather short timescale of $\sim 1500 r_g/c$. 
It will be important to assess which (if any) of these speculations is correct. 
{After we posted this work on the archives, \citet{2019arXiv190702079C} showed in a toroidal-field simulation of a compact, neutrino-cooled merger remnant accretion disc 
that dynamo can generate \emph{and} retain magnetic flux loops of alternating polarity, leading to striped jets (see also \citealt{2015MNRAS.446L..61P}).
The difference with our work can potentially emerge due to neutrino cooling and/or smaller disc size making their accretion flow more tightly bound and conducive to retaining the alternating-polarity dynamo-generated loops instead of expelling most of them in an outflow, as seen in Fig.~\ref{fig:contourplot}(a).}

A robust large-scale poloidal flux dynamo can help us understand the prevalence of jets across a wide range of astrophysical systems. Even though typical jet-producing accretion discs are thick near the BH, they may be thin at large radii \citep{Esin1997}. Thin discs are thought to be incapable of efficiently transporting large-scale poloidal magnetic flux from the ambient medium to the BH \citep{Lubow1994,Giulet2012,Giulet2013}, decreasing the prospects for jet formation. However, 
if the outer thin disc can transport even just a very weak poloidal magnetic flux, dynamo action in the inner, thick disc could amplify the magnetic flux in situ to levels sufficient for forming jets. In systems such as the jetted TDE Swift J1644+57 the stellar magnetic flux falls several orders of magnitude short of that necessary to power the observed jet \citep{Tchekhovskoy2014,2014MNRAS.445.3919K}. The rapid dynamo action in this work may amplify the available magnetic flux, explaining their observed jet power.

\section{Acknowledgments}

We thank P.~Bhat and P.~Dhang for discussions. This research was enabled by NSF PRAC awards 1615281, OAC-1811605 at the
Blue Waters computing project (for the \hammer{} simulation) and by the NASA High-End Computing (HEC) Program through the NASA Advanced Supercomputing (NAS) Division at Ames Research Center (for the \harm{} simulation). ML was supported by the NWO Spinoza
Prize (PI M.B.M. van der Klis). This work was supported in part by NSF
grant AST-1815304 and NASA grant 80NSSC18K0565 (AT), NSF grants AST 13-33612, AST 1715054, Chandra theory grant TM7-18006X from the Smithsonian Institution, and a Simons Investigator award from the Simons Foundation (EQ).

\section{Supporting Information}
Supplementary data are available at MNRAS online and  \href{https://www.youtube.com/channel/UCB2BslhkPMTkIicfoIBG0mQ}{here}.

\vspace{0.1cm}
\noindent Simulation data is available upon request from AT at
\href {mailto:atchekho@northwestern.edu}{atchekho@northwestern.edu}

\label{sec:acks}

{\small
\bibliography{mybib,sasha,newbib}

\begin{thebibliography}{}
\makeatletter
\relax
\def\mn@urlcharsother{\let\do\@makeother \do\$\do\&\do\#\do\^\do\_\do\%\do\~}
\def\mn@doi{\begingroup\mn@urlcharsother \@ifnextchar [ {\mn@doi@}
  {\mn@doi@[]}}
\def\mn@doi@[#1]#2{\def\@tempa{#1}\ifx\@tempa\@empty \href
  {http://dx.doi.org/#2} {doi:#2}\else \href {http://dx.doi.org/#2} {#1}\fi
  \endgroup}
\def\mn@eprint#1#2{\mn@eprint@#1:#2::\@nil}
\def\mn@eprint@arXiv#1{\href {http://arxiv.org/abs/#1} {{\tt arXiv:#1}}}
\def\mn@eprint@dblp#1{\href {http://dblp.uni-trier.de/rec/bibtex/#1.xml}
  {dblp:#1}}
\def\mn@eprint@#1:#2:#3:#4\@nil{\def\@tempa {#1}\def\@tempb {#2}\def\@tempc
  {#3}\ifx \@tempc \@empty \let \@tempc \@tempb \let \@tempb \@tempa \fi \ifx
  \@tempb \@empty \def\@tempb {arXiv}\fi \@ifundefined
  {mn@eprint@\@tempb}{\@tempb:\@tempc}{\expandafter \expandafter \csname
  mn@eprint@\@tempb\endcsname \expandafter{\@tempc}}}

\bibitem[\protect\citeauthoryear{{Bai} \& {Stone}}{{Bai} \&
  {Stone}}{2013}]{Bai2012}
{Bai} X.-N.,  {Stone} J.~M.,  2013, \mn@doi [\apj]
  {10.1088/0004-637X/767/1/30}, \href
  {http://adsabs.harvard.edu/abs/2013ApJ...767...30B} {767, 30}

\bibitem[\protect\citeauthoryear{{Balbus} \& {Hawley}}{{Balbus} \&
  {Hawley}}{1991}]{Balbus1991}
{Balbus} S.~A.,  {Hawley} J.~F.,  1991, \mn@doi [\apj] {10.1086/170270}, \href
  {http://adsabs.harvard.edu/abs/1991ApJ...376..214B} {376, 214}

\bibitem[\protect\citeauthoryear{{Beckwith}, {Hawley}  \& {Krolik}}{{Beckwith}
  et~al.}{2008}]{2008ApJ...678.1180B}
{Beckwith} K.,  {Hawley} J.~F.,   {Krolik} J.~H.,  2008, \mn@doi [\apj]
  {10.1086/533492}, \href {http://adsabs.harvard.edu/abs/2008ApJ...678.1180B}
  {678, 1180}

\bibitem[\protect\citeauthoryear{{Beckwith}, {Armitage}  \& {Simon}}{{Beckwith}
  et~al.}{2011}]{Beckwith2011}
{Beckwith} K.,  {Armitage} P.~J.,   {Simon} J.~B.,  2011, \mn@doi [\mnras]
  {10.1111/j.1365-2966.2011.19043.x}, \href
  {http://adsabs.harvard.edu/abs/2011MNRAS.416..361B} {416, 361}

\bibitem[\protect\citeauthoryear{{Begelman} \& {Pringle}}{{Begelman} \&
  {Pringle}}{2007}]{Begelman2007}
{Begelman} M.~C.,  {Pringle} J.~E.,  2007, \mn@doi [\mnras]
  {10.1111/j.1365-2966.2006.11372.x}, \href
  {https://ui.adsabs.harvard.edu/\#abs/2007MNRAS.375.1070B} {375, 1070}

\bibitem[\protect\citeauthoryear{{Bhat}, {Ebrahimi}  \& {Blackman}}{{Bhat}
  et~al.}{2016}]{2016MNRAS.462..818B}
{Bhat} P.,  {Ebrahimi} F.,   {Blackman} E.~G.,  2016, \mn@doi [\mnras]
  {10.1093/mnras/stw1619}, \href
  {http://adsabs.harvard.edu/abs/2016MNRAS.462..818B} {462, 818}

\bibitem[\protect\citeauthoryear{{Blandford} \& {Znajek}}{{Blandford} \&
  {Znajek}}{1977}]{bz77}
{Blandford} R.~D.,  {Znajek} R.~L.,  1977, \mn@doi [\mnras]
  {10.1093/mnras/179.3.433}, \href
  {http://adsabs.harvard.edu/abs/1977MNRAS.179..433B} {179, 433}

\bibitem[\protect\citeauthoryear{{Brandenburg}, {Nordlund}, {Stein}  \&
  {Torkelsson}}{{Brandenburg} et~al.}{1995}]{Brandenburg1995}
{Brandenburg} A.,  {Nordlund} A.,  {Stein} R.~F.,   {Torkelsson} U.,  1995,
  \mn@doi [\apj] {10.1086/175831}, \href
  {http://adsabs.harvard.edu/abs/1995ApJ...446..741B} {446, 741}

\bibitem[\protect\citeauthoryear{{Bromberg} \& {Tchekhovskoy}}{{Bromberg} \&
  {Tchekhovskoy}}{2016}]{Bromberg2016}
{Bromberg} O.,  {Tchekhovskoy} A.,  2016, \mn@doi [\mnras]
  {10.1093/mnras/stv2591}, \href
  {http://adsabs.harvard.edu/abs/2016MNRAS.456.1739B} {456, 1739}

\bibitem[\protect\citeauthoryear{{Chakrabarti}}{{Chakrabarti}}{1985}]{1985ApJ...288....1C}
{Chakrabarti} S.~K.,  1985, \mn@doi [\apj] {10.1086/162755}, \href
  {http://adsabs.harvard.edu/abs/1985ApJ...288....1C} {288, 1}

\bibitem[\protect\citeauthoryear{{Chatterjee}, {Liska}, {Tchekhovskoy}  \&
  {Markoff}}{{Chatterjee} et~al.}{2019}]{2019arXiv190403243C}
{Chatterjee} K.,  {Liska} M.,  {Tchekhovskoy} A.,   {Markoff} S.~B.,  2019,
  arXiv e-prints, \href {https://ui.adsabs.harvard.edu/abs/2019arXiv190403243C}
  {p. arXiv:1904.03243}

\bibitem[\protect\citeauthoryear{{Christie}, {Lalakos}, {Tchekhovskoy},
  {Fern{\'a}ndez}, {Foucart}, {Quataert}  \& {Kasen}}{{Christie}
  et~al.}{2019}]{2019arXiv190702079C}
{Christie} I.~M.,  {Lalakos} A.,  {Tchekhovskoy} A.,  {Fern{\'a}ndez} R.,
  {Foucart} F.,  {Quataert} E.,   {Kasen} D.,  2019, arXiv e-prints, \href
  {https://ui.adsabs.harvard.edu/abs/2019arXiv190702079C} {p. arXiv:1907.02079}

\bibitem[\protect\citeauthoryear{{Colella} \& {Woodward}}{{Colella} \&
  {Woodward}}{1984}]{1984JCoPh..54..174C}
{Colella} P.,  {Woodward} P.~R.,  1984, \mn@doi [JCP]
  {10.1016/0021-9991(84)90143-8}, \href
  {http://adsabs.harvard.edu/abs/1984JCoPh..54..174C} {54, 174}

\bibitem[\protect\citeauthoryear{{Davis}, {Stone}  \& {Pessah}}{{Davis}
  et~al.}{2010}]{Davis2010}
{Davis} S.~W.,  {Stone} J.~M.,   {Pessah} M.~E.,  2010, \mn@doi [\apj]
  {10.1088/0004-637X/713/1/52}, \href
  {http://adsabs.harvard.edu/abs/2010ApJ...713...52D} {713, 52}

\bibitem[\protect\citeauthoryear{{De Villiers} \& {Hawley}}{{De Villiers} \&
  {Hawley}}{2003}]{2003ApJ...589..458D}
{De Villiers} J.-P.,  {Hawley} J.~F.,  2003, \mn@doi [\apj] {10.1086/373949},
  \href {http://adsabs.harvard.edu/abs/2003ApJ...589..458D} {589, 458}

\bibitem[\protect\citeauthoryear{{Dhang} \& {Sharma}}{{Dhang} \&
  {Sharma}}{2019}]{2019MNRAS.482..848D}
{Dhang} P.,  {Sharma} P.,  2019, \mn@doi [\mnras] {10.1093/mnras/sty2692},
  \href {https://ui.adsabs.harvard.edu/abs/2019MNRAS.482..848D} {482, 848}

\bibitem[\protect\citeauthoryear{{Esin}, {McClintock}  \& {Narayan}}{{Esin}
  et~al.}{1997}]{Esin1997}
{Esin} A.~A.,  {McClintock} J.~E.,   {Narayan} R.,  1997, \mn@doi [\apj]
  {10.1086/304829}, \href {http://adsabs.harvard.edu/abs/1997ApJ...489..865E}
  {489, 865}

\bibitem[\protect\citeauthoryear{{Fabian}}{{Fabian}}{2012}]{Fabian2012}
{Fabian} A.~C.,  2012, \mn@doi [\araa] {10.1146/annurev-astro-081811-125521},
  \href {http://adsabs.harvard.edu/abs/2012ARA%26A..50..455F} {50, 455}

\bibitem[\protect\citeauthoryear{{Flock}, {Dzyurkevich}, {Klahr}, {Turner}  \&
  {Henning}}{{Flock} et~al.}{2012}]{Flock2012}
{Flock} M.,  {Dzyurkevich} N.,  {Klahr} H.,  {Turner} N.,   {Henning} T.,
  2012, \mn@doi [\apj] {10.1088/0004-637X/744/2/144}, \href
  {http://adsabs.harvard.edu/abs/2012ApJ...744..144F} {744, 144}

\bibitem[\protect\citeauthoryear{{Fragile} \& {Sadowski}}{{Fragile} \&
  {Sadowski}}{2017}]{Fragile2017}
{Fragile} P.~C.,  {Sadowski} A.,  2017, \mn@doi [\mnras]
  {10.1093/mnras/stx274}, \href
  {https://ui.adsabs.harvard.edu/abs/2017MNRAS.467.1838F} {467, 1838}

\bibitem[\protect\citeauthoryear{{Gammie}, {McKinney}  \& {T{\'o}th}}{{Gammie}
  et~al.}{2003}]{Gammie2003}
{Gammie} C.~F.,  {McKinney} J.~C.,   {T{\'o}th} G.,  2003, \mn@doi [\apj]
  {10.1086/374594}, \href {http://adsabs.harvard.edu/abs/2003ApJ...589..444G}
  {589, 444}

\bibitem[\protect\citeauthoryear{{Ghisellini}, {Tavecchio}, {Maraschi},
  {Celotti}  \& {Sbarrato}}{{Ghisellini} et~al.}{2014}]{Ghissellini2014}
{Ghisellini} G.,  {Tavecchio} F.,  {Maraschi} L.,  {Celotti} A.,   {Sbarrato}
  T.,  2014, \mn@doi [\nat] {10.1038/nature13856}, \href
  {http://adsabs.harvard.edu/abs/2014Natur.515..376G} {515, 376}

\bibitem[\protect\citeauthoryear{{Guilet} \& {Ogilvie}}{{Guilet} \&
  {Ogilvie}}{2012}]{Giulet2012}
{Guilet} J.,  {Ogilvie} G.~I.,  2012, \mn@doi [\mnras]
  {10.1111/j.1365-2966.2012.21361.x}, \href
  {http://adsabs.harvard.edu/abs/2012MNRAS.424.2097G} {424, 2097}

\bibitem[\protect\citeauthoryear{{Guilet} \& {Ogilvie}}{{Guilet} \&
  {Ogilvie}}{2013}]{Giulet2013}
{Guilet} J.,  {Ogilvie} G.~I.,  2013, \mn@doi [\mnras] {10.1093/mnras/sts551},
  \href {http://adsabs.harvard.edu/abs/2013MNRAS.430..822G} {430, 822}

\bibitem[\protect\citeauthoryear{{Hawley} \& {Krolik}}{{Hawley} \&
  {Krolik}}{2006}]{2006ApJ...641..103H}
{Hawley} J.~F.,  {Krolik} J.~H.,  2006, \mn@doi [\apj] {10.1086/500385}, \href
  {http://adsabs.harvard.edu/abs/2006ApJ...641..103H} {641, 103}

\bibitem[\protect\citeauthoryear{{Hogg} \& {Reynolds}}{{Hogg} \&
  {Reynolds}}{2018}]{2018ApJ...861...24H}
{Hogg} J.~D.,  {Reynolds} C.~S.,  2018, \mn@doi [\apj]
  {10.3847/1538-4357/aac439}, \href
  {http://adsabs.harvard.edu/abs/2018ApJ...861...24H} {861, 24}

\bibitem[\protect\citeauthoryear{{Jiang}, {Stone}  \& {Davis}}{{Jiang}
  et~al.}{2017}]{Jiang2017}
{Jiang} Y.-F.,  {Stone} J.,   {Davis} S.~W.,  2017, ArXiv:1709.02845, \href
  {http://adsabs.harvard.edu/abs/2017arXiv170902845J} {}

\bibitem[\protect\citeauthoryear{{Jiang}, {Blaes}, {Stone}  \& {Davis}}{{Jiang}
  et~al.}{2019}]{2019arXiv190401674J}
{Jiang} Y.-F.,  {Blaes} O.,  {Stone} J.,   {Davis} S.~W.,  2019, arXiv
  e-prints, \href {https://ui.adsabs.harvard.edu/abs/2019arXiv190401674J} {p.
  arXiv:1904.01674}

\bibitem[\protect\citeauthoryear{{Kelley}, {Tchekhovskoy}  \&
  {Narayan}}{{Kelley} et~al.}{2014}]{2014MNRAS.445.3919K}
{Kelley} L.~Z.,  {Tchekhovskoy} A.,   {Narayan} R.,  2014, \mn@doi [\mnras]
  {10.1093/mnras/stu2041}, \href
  {http://adsabs.harvard.edu/abs/2014MNRAS.445.3919K} {445, 3919}

\bibitem[\protect\citeauthoryear{{Komissarov}}{{Komissarov}}{2001}]{2001MNRAS.326L..41K}
{Komissarov} S.~S.,  2001, \mn@doi [\mnras] {10.1046/j.1365-8711.2001.04863.x},
  \href {http://adsabs.harvard.edu/abs/2001MNRAS.326L..41K} {326, L41}

\bibitem[\protect\citeauthoryear{{Lesur} \& {Ogilvie}}{{Lesur} \&
  {Ogilvie}}{2008}]{2008A&A...488..451L}
{Lesur} G.,  {Ogilvie} G.~I.,  2008, \mn@doi [\aap]
  {10.1051/0004-6361:200810152}, \href
  {http://adsabs.harvard.edu/abs/2008A%26A...488..451L} {488, 451}

\bibitem[\protect\citeauthoryear{{Liska}, {Hesp}, {Tchekhovskoy}, {Ingram},
  {van der Klis}  \& {Markoff}}{{Liska} et~al.}{2018}]{Liska2018A}
{Liska} M.,  {Hesp} C.,  {Tchekhovskoy} A.,  {Ingram} A.,  {van der Klis} M.,
  {Markoff} S.,  2018, \mn@doi [\mnras] {10.1093/mnrasl/slx174}, \href
  {http://adsabs.harvard.edu/abs/2018MNRAS.474L..81L} {474, L81}

\bibitem[\protect\citeauthoryear{{Liska} et~al.,}{{Liska}
  et~al.}{2019a}]{2019arXiv191210192L}
{Liska} M.,  et~al., 2019a, arXiv e-prints, \href
  {https://ui.adsabs.harvard.edu/abs/2019arXiv191210192L} {p. arXiv:1912.10192}

\bibitem[\protect\citeauthoryear{{Liska}, {Tchekhovskoy}, {Ingram}  \& {van der
  Klis}}{{Liska} et~al.}{2019b}]{Liska2018C}
{Liska} M.,  {Tchekhovskoy} A.,  {Ingram} A.,   {van der Klis} M.,  2019b,
  \mn@doi [\mnras] {10.1093/mnras/stz834}, \href
  {https://ui.adsabs.harvard.edu/abs/2019MNRAS.487..550L} {487, 550}

\bibitem[\protect\citeauthoryear{Lubow, Papaloizou  \& Pringle}{Lubow
  et~al.}{1994}]{Lubow1994}
Lubow S.~H.,  Papaloizou J. C.~B.,   Pringle J.~E.,  1994, \mn@doi [MNRAS]
  {10.1093/mnras/267.2.235}, 267, 235

\bibitem[\protect\citeauthoryear{{McKinney}}{{McKinney}}{2006}]{Mckinney2006}
{McKinney} J.~C.,  2006, \mn@doi [\mnras] {10.1111/j.1365-2966.2006.10256.x},
  \href {http://adsabs.harvard.edu/abs/2006MNRAS.368.1561M} {368, 1561}

\bibitem[\protect\citeauthoryear{{McKinney}, {Tchekhovskoy}  \&
  {Blandford}}{{McKinney} et~al.}{2012}]{Mckinney2012}
{McKinney} J.~C.,  {Tchekhovskoy} A.,   {Blandford} R.~D.,  2012, \mn@doi
  [\mnras] {10.1111/j.1365-2966.2012.21074.x}, \href
  {http://adsabs.harvard.edu/abs/2012MNRAS.423.3083M} {423, 3083}

\bibitem[\protect\citeauthoryear{{Mertens}, {Lobanov}, {Walker}  \&
  {Hardee}}{{Mertens} et~al.}{2016}]{2016A&A...595A..54M}
{Mertens} F.,  {Lobanov} A.~P.,  {Walker} R.~C.,   {Hardee} P.~E.,  2016,
  \mn@doi [\aap] {10.1051/0004-6361/201628829}, \href
  {http://adsabs.harvard.edu/abs/2016A%26A...595A..54M} {595, A54}

\bibitem[\protect\citeauthoryear{{Moffatt}}{{Moffatt}}{1978}]{1978mfge.book.....M}
{Moffatt} H.~K.,  1978, {Magnetic field generation in electrically conducting
  fluids}.
Cambridge, England, Cambridge University Press.~353 p.

\bibitem[\protect\citeauthoryear{{Na\-ra\-yan}, {Igumenshchev}  \&
  {Abramowicz}}{{Na\-ra\-yan} et~al.}{2003}]{Narayan2003}
{Na\-ra\-yan} R.,  {Igumenshchev} I.~V.,   {Abramowicz} M.~A.,  2003, \mn@doi
  [\pasj] {10.1093/pasj/55.6.L69}, \href
  {http://adsabs.harvard.edu/abs/2003PASJ...55L..69N} {55, L69}

\bibitem[\protect\citeauthoryear{{Nakamura} \& {Asada}}{{Nakamura} \&
  {Asada}}{2013}]{2013ApJ...775..118N}
{Nakamura} M.,  {Asada} K.,  2013, \mn@doi [\apj]
  {10.1088/0004-637X/775/2/118}, \href
  {http://adsabs.harvard.edu/abs/2013ApJ...775..118N} {775, 118}

\bibitem[\protect\citeauthoryear{{Narayan}, {S{\c a}dowski}, {Penna}  \&
  {Kulkarni}}{{Narayan} et~al.}{2012}]{2012MNRAS.426.3241N}
{Narayan} R.,  {S{\c a}dowski} A.,  {Penna} R.~F.,   {Kulkarni} A.~K.,  2012,
  \mn@doi [\mnras] {10.1111/j.1365-2966.2012.22002.x}, \href
  {http://adsabs.harvard.edu/abs/2012MNRAS.426.3241N} {426, 3241}

\bibitem[\protect\citeauthoryear{{Nauman} \& {Blackman}}{{Nauman} \&
  {Blackman}}{2015}]{2015MNRAS.446.2102N}
{Nauman} F.,  {Blackman} E.~G.,  2015, \mn@doi [\mnras]
  {10.1093/mnras/stu2226}, \href
  {https://ui.adsabs.harvard.edu/abs/2015MNRAS.446.2102N} {446, 2102}

\bibitem[\protect\citeauthoryear{{Nemmen} \& {Tchekhovskoy}}{{Nemmen} \&
  {Tchekhovskoy}}{2015}]{2015MNRAS.449..316N}
{Nemmen} R.~S.,  {Tchekhovskoy} A.,  2015, \mn@doi [\mnras]
  {10.1093/mnras/stv260}, \href
  {http://adsabs.harvard.edu/abs/2015MNRAS.449..316N} {449, 316}

\bibitem[\protect\citeauthoryear{{Noble}, {Krolik}  \& {Hawley}}{{Noble}
  et~al.}{2009}]{Noble2009}
{Noble} S.~C.,  {Krolik} J.~H.,   {Hawley} J.~F.,  2009, \mn@doi [\apj]
  {10.1088/0004-637X/692/1/411}, \href
  {http://adsabs.harvard.edu/abs/2009ApJ...692..411N} {692, 411}

\bibitem[\protect\citeauthoryear{{O'Neill}, {Reynolds}, {Miller}  \&
  {Sorathia}}{{O'Neill} et~al.}{2011}]{Oneill2011}
{O'Neill} S.~M.,  {Reynolds} C.~S.,  {Miller} M.~C.,   {Sorathia} K.~A.,  2011,
  \mn@doi [\apj] {10.1088/0004-637X/736/2/107}, \href
  {http://adsabs.harvard.edu/abs/2011ApJ...736..107O} {736, 107}

\bibitem[\protect\citeauthoryear{{Parfrey}, {Giannios}  \&
  {Beloborodov}}{{Parfrey} et~al.}{2015}]{2015MNRAS.446L..61P}
{Parfrey} K.,  {Giannios} D.,   {Beloborodov} A.~M.,  2015, \mn@doi [\mnras]
  {10.1093/mnrasl/slu162}, \href
  {https://ui.adsabs.harvard.edu/abs/2015MNRAS.446L..61P} {446, L61}

\bibitem[\protect\citeauthoryear{{Parker}}{{Parker}}{1955}]{1955ApJ...122..293P}
{Parker} E.~N.,  1955, \mn@doi [\apj] {10.1086/146087}, \href
  {http://adsabs.harvard.edu/abs/1955ApJ...122..293P} {122, 293}

\bibitem[\protect\citeauthoryear{{Penna}, {McKinney}, {Narayan},
  {Tchekhovskoy}, {Shafee}  \& {McClintock}}{{Penna} et~al.}{2010}]{Penna2010}
{Penna} R.~F.,  {McKinney} J.~C.,  {Narayan} R.,  {Tchekhovskoy} A.,  {Shafee}
  R.,   {McClintock} J.~E.,  2010, \mn@doi [\mnras]
  {10.1111/j.1365-2966.2010.17170.x}, \href
  {http://adsabs.harvard.edu/abs/2010MNRAS.408..752P} {408, 752}

\bibitem[\protect\citeauthoryear{{Pessah}, {Chan}  \& {Psaltis}}{{Pessah}
  et~al.}{2007}]{2007ApJ...668L..51P}
{Pessah} M.~E.,  {Chan} C.-k.,   {Psaltis} D.,  2007, \mn@doi [\apjl]
  {10.1086/522585}, \href {http://adsabs.harvard.edu/abs/2007ApJ...668L..51P}
  {668, L51}

\bibitem[\protect\citeauthoryear{{Prieto}, {Fern{\'a}ndez-Ontiveros},
  {Markoff}, {Espada}  \& {Gonz{\'a}lez-Mart{\'{\i}}n}}{{Prieto}
  et~al.}{2016}]{Prieto2016}
{Prieto} M.~A.,  {Fern{\'a}ndez-Ontiveros} J.~A.,  {Markoff} S.,  {Espada} D.,
   {Gonz{\'a}lez-Mart{\'{\i}}n} O.,  2016, \mn@doi [\mnras]
  {10.1093/mnras/stw166}, \href
  {http://adsabs.harvard.edu/abs/2016MNRAS.457.3801P} {457, 3801}

\bibitem[\protect\citeauthoryear{{Ressler}, {Tchekhovskoy}, {Quataert}  \&
  {Gammie}}{{Ressler} et~al.}{2017}]{Ressler2017}
{Ressler} S.~M.,  {Tchekhovskoy} A.,  {Quataert} E.,   {Gammie} C.~F.,  2017,
  \mn@doi [\mnras] {10.1093/mnras/stx364}, \href
  {http://adsabs.harvard.edu/abs/2017MNRAS.467.3604R} {467, 3604}

\bibitem[\protect\citeauthoryear{{Rothstein} \& {Lovelace}}{{Rothstein} \&
  {Lovelace}}{2008}]{Rothstein2008}
{Rothstein} D.~M.,  {Lovelace} R.~V.~E.,  2008, \mn@doi [\apj]
  {10.1086/529128}, \href {http://adsabs.harvard.edu/abs/2008ApJ...677.1221R}
  {677, 1221}

\bibitem[\protect\citeauthoryear{{Ryan}, {Gammie}, {Fromang}  \&
  {Kestener}}{{Ryan} et~al.}{2017}]{2017ApJ...840....6R}
{Ryan} B.~R.,  {Gammie} C.~F.,  {Fromang} S.,   {Kestener} P.,  2017, \mn@doi
  [\apj] {10.3847/1538-4357/aa6a52}, \href
  {http://adsabs.harvard.edu/abs/2017ApJ...840....6R} {840, 6}

\bibitem[\protect\citeauthoryear{{Salvesen}, {Simon}, {Armitage}  \&
  {Begelman}}{{Salvesen} et~al.}{2016a}]{Salvesen2016}
{Salvesen} G.,  {Simon} J.~B.,  {Armitage} P.~J.,   {Begelman} M.~C.,  2016a,
  \mn@doi [\mnras] {10.1093/mnras/stw029}, \href
  {http://adsabs.harvard.edu/abs/2016MNRAS.457..857S} {457, 857}

\bibitem[\protect\citeauthoryear{{Salvesen}, {Armitage}, {Simon}  \&
  {Begelman}}{{Salvesen} et~al.}{2016b}]{2016MNRAS.460.3488S}
{Salvesen} G.,  {Armitage} P.~J.,  {Simon} J.~B.,   {Begelman} M.~C.,  2016b,
  \mn@doi [\mnras] {10.1093/mnras/stw1231}, \href
  {http://adsabs.harvard.edu/abs/2016MNRAS.460.3488S} {460, 3488}

\bibitem[\protect\citeauthoryear{{S{\c{a}}dowski}}{{S{\c{a}}dowski}}{2016}]{2016MNRAS.459.4397S}
{S{\c{a}}dowski} A.,  2016, \mn@doi [\mnras] {10.1093/mnras/stw913}, \href
  {https://ui.adsabs.harvard.edu/abs/2016MNRAS.459.4397S} {459, 4397}

\bibitem[\protect\citeauthoryear{{Shafee}, {McKinney}, {Narayan},
  {Tchekhovskoy}, {Gammie}  \& {McClintock}}{{Shafee}
  et~al.}{2008}]{2008ApJ...687L..25S}
{Shafee} R.,  {McKinney} J.~C.,  {Narayan} R.,  {Tchekhovskoy} A.,  {Gammie}
  C.~F.,   {McClintock} J.~E.,  2008, \mn@doi [\apjl] {10.1086/593148}, \href
  {http://adsabs.harvard.edu/abs/2008ApJ...687L..25S} {687, L25}

\bibitem[\protect\citeauthoryear{{Shi}, {Krolik}  \& {Hirose}}{{Shi}
  et~al.}{2010}]{Shi2010}
{Shi} J.,  {Krolik} J.~H.,   {Hirose} S.,  2010, \mn@doi [\apj]
  {10.1088/0004-637X/708/2/1716}, \href
  {http://adsabs.harvard.edu/abs/2010ApJ...708.1716S} {708, 1716}

\bibitem[\protect\citeauthoryear{{Shi}, {Stone}  \& {Huang}}{{Shi}
  et~al.}{2016}]{2016MNRAS.456.2273S}
{Shi} J.-M.,  {Stone} J.~M.,   {Huang} C.~X.,  2016, \mn@doi [\mnras]
  {10.1093/mnras/stv2815}, \href
  {http://adsabs.harvard.edu/abs/2016MNRAS.456.2273S} {456, 2273}

\bibitem[\protect\citeauthoryear{{Siegel} \& {Metzger}}{{Siegel} \&
  {Metzger}}{2018}]{Siegel2017}
{Siegel} D.~M.,  {Metzger} B.~D.,  2018, \mn@doi [\apj]
  {10.3847/1538-4357/aabaec}, \href
  {http://adsabs.harvard.edu/abs/2018ApJ...858...52S} {858, 52}

\bibitem[\protect\citeauthoryear{{Simon}, {Hawley}  \& {Beckwith}}{{Simon}
  et~al.}{2011}]{Simon2011}
{Simon} J.~B.,  {Hawley} J.~F.,   {Beckwith} K.,  2011, \mn@doi [\apj]
  {10.1088/0004-637X/730/2/94}, \href
  {http://adsabs.harvard.edu/abs/2011ApJ...730...94S} {730, 94}

\bibitem[\protect\citeauthoryear{{Simon}, {Beckwith}  \& {Armitage}}{{Simon}
  et~al.}{2012}]{Simon2012}
{Simon} J.~B.,  {Beckwith} K.,   {Armitage} P.~J.,  2012, \mn@doi [\mnras]
  {10.1111/j.1365-2966.2012.20835.x}, \href
  {http://adsabs.harvard.edu/abs/2012MNRAS.422.2685S} {422, 2685}

\bibitem[\protect\citeauthoryear{{Stone}, {Hawley}, {Gammie}  \&
  {Balbus}}{{Stone} et~al.}{1996}]{1996ApJ...463..656S}
{Stone} J.~M.,  {Hawley} J.~F.,  {Gammie} C.~F.,   {Balbus} S.~A.,  1996,
  \mn@doi [\apj] {10.1086/177280}, \href
  {http://adsabs.harvard.edu/abs/1996ApJ...463..656S} {463, 656}

\bibitem[\protect\citeauthoryear{{Tchekhovskoy} \& {Bromberg}}{{Tchekhovskoy}
  \& {Bromberg}}{2016}]{Tchekhovskoy2016}
{Tchekhovskoy} A.,  {Bromberg} O.,  2016, \mn@doi [\mnras]
  {10.1093/mnrasl/slw064}, \href
  {http://adsabs.harvard.edu/abs/2016MNRAS.461L..46T} {461, L46}

\bibitem[\protect\citeauthoryear{{Tchekhovskoy} \& {McKinney}}{{Tchekhovskoy}
  \& {McKinney}}{2012}]{Tchekhovskoy2012}
{Tchekhovskoy} A.,  {McKinney} J.~C.,  2012, \mn@doi [\mnras]
  {10.1111/j.1745-3933.2012.01256.x}, \href
  {http://adsabs.harvard.edu/abs/2012MNRAS.423L..55T} {423, L55}

\bibitem[\protect\citeauthoryear{{Tchekhovskoy}, {McKinney}  \&
  {Narayan}}{{Tchekhovskoy} et~al.}{2008}]{Tchekhovskoy2008}
{Tchekhovskoy} A.,  {McKinney} J.~C.,   {Narayan} R.,  2008, \mn@doi [\mnras]
  {10.1111/j.1365-2966.2008.13425.x}, \href
  {http://adsabs.harvard.edu/abs/2008MNRAS.388..551T} {388, 551}

\bibitem[\protect\citeauthoryear{{Tchekhovskoy}, {Narayan}  \&
  {McKinney}}{{Tchekhovskoy} et~al.}{2010}]{tchekhovskoy10}
{Tchekhovskoy} A.,  {Narayan} R.,   {McKinney} J.~C.,  2010, \mn@doi [\apj]
  {10.1088/0004-637X/711/1/50}, \href
  {http://adsabs.harvard.edu/abs/2010ApJ...711...50T} {711, 50}

\bibitem[\protect\citeauthoryear{{Tchekhovskoy}, {Narayan}  \&
  {McKinney}}{{Tchekhovskoy} et~al.}{2011}]{Tchekhovskoy2011}
{Tchekhovskoy} A.,  {Narayan} R.,   {McKinney} J.~C.,  2011, \mn@doi [\mnras]
  {10.1111/j.1745-3933.2011.01147.x}, \href
  {http://adsabs.harvard.edu/abs/2011MNRAS.418L..79T} {418, L79}

\bibitem[\protect\citeauthoryear{{Tchekhovskoy}, {Metzger}, {Giannios}  \&
  {Kelley}}{{Tchekhovskoy} et~al.}{2014}]{Tchekhovskoy2014}
{Tchekhovskoy} A.,  {Metzger} B.~D.,  {Giannios} D.,   {Kelley} L.~Z.,  2014,
  \mn@doi [\mnras] {10.1093/mnras/stt2085}, \href
  {http://adsabs.harvard.edu/abs/2014MNRAS.437.2744T} {437, 2744}

\bibitem[\protect\citeauthoryear{{White}, {Stone}  \& {Quataert}}{{White}
  et~al.}{2019}]{2019ApJ...874..168W}
{White} C.~J.,  {Stone} J.~M.,   {Quataert} E.,  2019, \mn@doi [\apj]
  {10.3847/1538-4357/ab0c0c}, \href
  {https://ui.adsabs.harvard.edu/abs/2019ApJ...874..168W} {874, 168}

\makeatother
\end{thebibliography}
\bibliographystyle{mnras}
}

\appendix
\section{Large-scale Jet Properties}
\label{sec:large-scale-jet}
\begin{figure}
\begin{center}
\includegraphics[width=0.8\columnwidth,trim=0mm 0mm 0mm
0mm,clip]{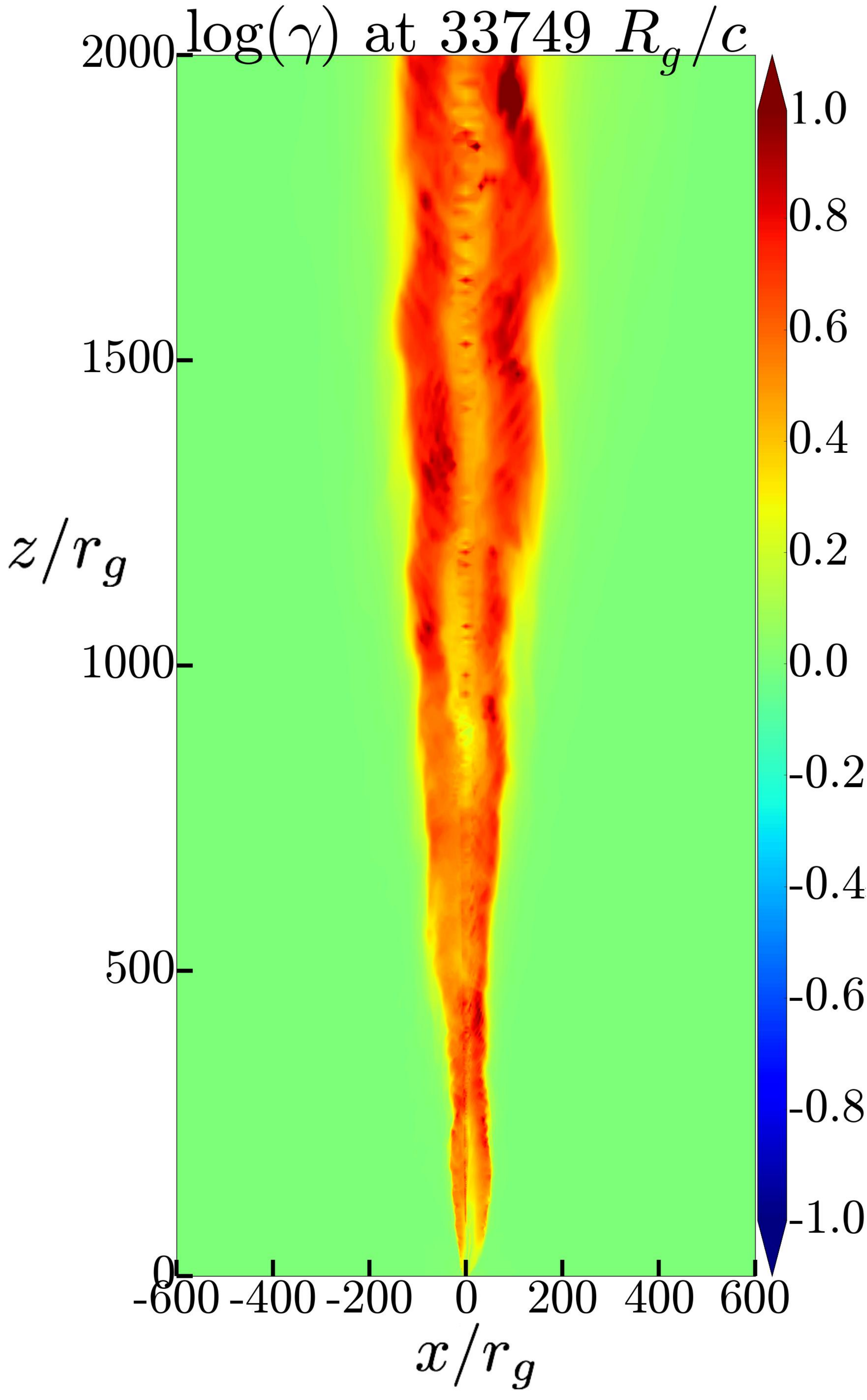}
\end{center}
\caption{Vertical slice through the Lorentz factor
    in a simulation snapshot of the upper relativistic jet extending out to $r=2000r_g$. While the jets show deviations from axisymmetry, they do not undergo significant kink of pinch instabilities and reach Lorentz factors of $\gamma \sim 10$. As the jets accelerate, they collimate into progressively smaller opening angles, which become as small as $\theta\sim0.08$ (see also Fig.~\ref{fig:jetplot}).}  \label{fig:jet_zoomout} \end{figure}

Fig.~\ref{fig:jet_zoomout} shows a vertical slice through the Lorentz factor of one of our jets. The jets have a fast sheath and a slow spine (see also \citealt{Tchekhovskoy2008}). While the jets are not axisymmetric, they remain largely stable and accelerate to Lorentz factors of $\gamma\sim10$ at distances of $z = 2000r_g$.

\section{Description of the Low-resolution HARM simulation}
\label{sec:descr-low-resol}
\begin{figure}
\begin{center}
\includegraphics[width=\columnwidth,trim=0mm 0mm 0mm 0mm,clip]{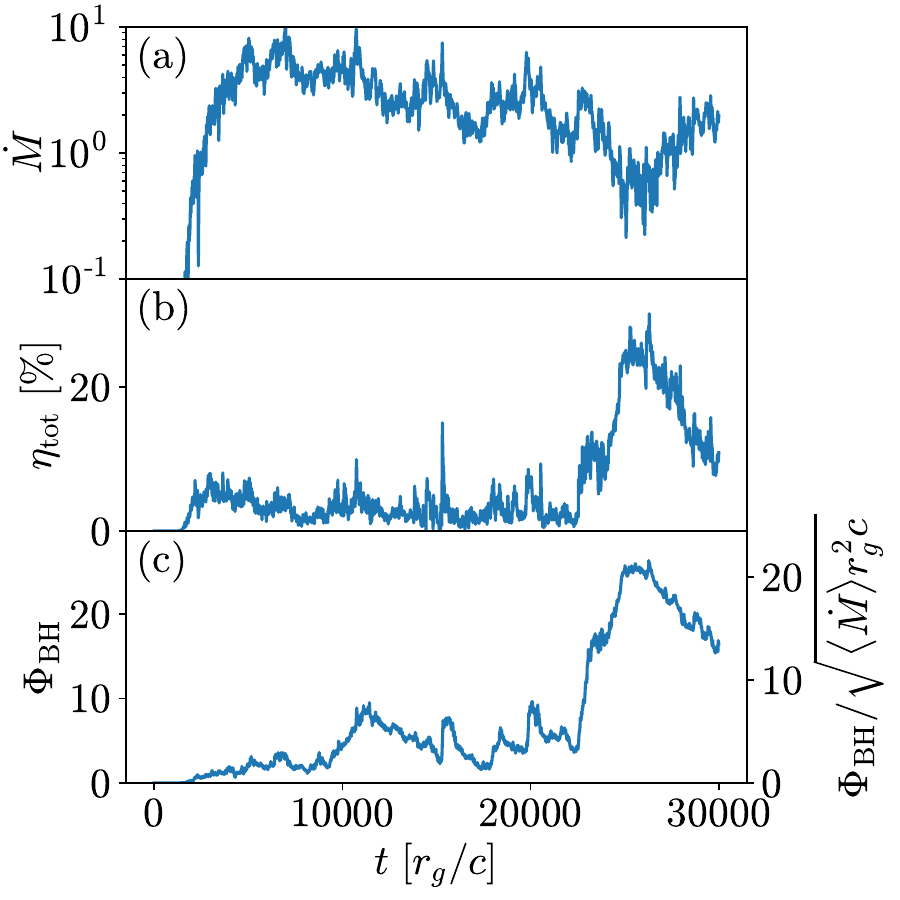}
\end{center}
\caption{Similar to Fig.~\ref{fig:timeplot} but for our low-resolution HARM simulation. While [panel (a)] the mass accretion rate shows a qualitatively similar behavior -- peak around $\dot M\approx6$ before declining -- as for our fiducial run, [panel (b)] the outflow energy efficiency reaches $\eta_{\rm tot}\simeq 25\%$ for this lower-resolution run, twice as low as for the fiducial model over the same time interval (see Fig.~\ref{fig:timeplot}b), and [panel (c)] the magnetic flux reaches $\Phi_{\rm BH}\approx 25$, also remains approximately twice as small as for the fiducial model (see Fig.~\ref{fig:timeplot}c). The magnetic flux normalized to the late-time-average mass accretion rate ($\langle \dot M\rangle\approx1.5$) reaches $\simeq20$ (right axis in panel (c)).}
\label{fig:harm_sim}
\end{figure}

We can glean the effect of changing numerical resolution on the large-scale poloidal magnetic flux dynamo by comparing to a lower-resolution \HARM{} simulation that showed the signs of large-scale poloidal magnetic flux dynamo for the first time. We carried it out using the \HARM{} code at a resolution of $288\times128\times128$, with the $\theta$-grid focused on the equator at small distances while at the same time collimating towards the polar regions to resolve the jets at large distances (see \citealt{Tchekhovskoy2011,Ressler2017}). This simulation used a reduced toroidal wedge of $\Delta\phi = \pi$, with periodic boundary conditions in $\varphi$. The focusing of the numerical grid towards the equatorial plane results in grid cell aspect ratio near the equatorial plane of $\Delta r:r\Delta\theta:r\Delta\phi \approx 3:1:4$, i.e., cells elongated in radial and toroidal directions. Figure~\ref{fig:harm_sim} shows the time-dependence of mass accretion rate, jet efficiency, and magnetic flux versus time. As you can see, by $t \sim 30,000r_g/c$, the \HARM{} simulation reaches outflow energy efficiency of $\eta \sim 25$\%, about half of that in the \hammer{} simulation. This suggests that large-scale poloidal magnetic flux dynamo is more vigorous at a higher resolution, full azimuthal extent of the grid, and/or near-unity aspect ratio of the cells within the accretion disk. We will determine the factors important for resolving the dynamo activity in future work.

\label{lastpage}
\end{document}